\documentclass{article}
\usepackage{bm}
\usepackage{amsmath}
\usepackage{balance}
\DeclareMathOperator{\pen}{pen}
\newcommand{\norm}[1]{\left\lVert#1\right\rVert}
\usepackage{url}
\usepackage{algorithm}
\usepackage{algorithmic}

\usepackage{hyperref}
\usepackage{epstopdf}
\usepackage{stackrel}
\usepackage{graphicx}
\usepackage[utf8]{inputenc}
\usepackage{amssymb}
\usepackage{amsthm}
\usepackage{amsmath, calc}
\usepackage[authoryear]{natbib}

\begin{document}


\title{selectBoost: a general algorithm to enhance the performance of variable selection methods in correlated datasets}
\author{Isma\" il Aouadi\,$^{\text{ 1,2,3,4,}}$, Nicolas Jung\,$^{\text{ 1,5}}$, Raphael Carapito\,$^{\text{ 1,2,3,4}}$,\\
 Laurent Vallat\,$^{\text{ 1,3,4,}\footnote{present address: IRFAC, INSERM UMR\_S1113, F\'{e}d\'{e}ration de M\'{e}decine Translationnelle de Strasbourg (FMTS), Universit\'{e} de Strasbourg, Strasbourg, France.}}$, Seiamak Bahram\,$^{\text{ 1,2,3,4}}$, Myriam Maumy-Bertrand\,$^{\text{ 5}}$ \\ and Fr\'{e}d\'{e}ric Bertrand\,$^{\text{ 5,}\footnote{$^\ast$To whom correspondence should be addressed}}$\\
$^{\text{\sf 1}}$ImmunoRhumatologie Mol\'{e}culaire, \\
INSERM UMR\_S 1109, LabEx TRANSPLANTEX, \\
Centre de Recherche d'Immunologie et d'H\'{e}matologie, \\
Facult\'{e} de M\'{e}decine, \\
F\'{e}d\'{e}ration de M\'{e}decine Translationnelle de Strasbourg (FMTS), \\
Universit\'{e} de Strasbourg, Strasbourg, France \\
$^{\text{\sf 2}}$Laboratoire International Associ\'{e} (LIA) INSERM, \\
Strasbourg (France) - Nagano (Japan), Strasbourg, France \\
$^{\text{\sf 3}}$F\'{e}d\'{e}ration Hospitalo-Universitaire \\
(FHU) OMICARE, Strasbourg, France \\
$^{\text{\sf 4}}$Laboratoire Central d'Immunologie, P\^{o}le de Biologie, \\
Nouvel H\^{o}pital Civil, H\^{o}pitaux Universitaires de Strasbourg, \\
Strasbourg, France \\
$^{\text{\sf 5}}$Institut de Recherche Math\'ematique Avanc\'ee, CNRS UMR 7501, \\
Labex IRMIA, Universit\'{e} de Strasbourg, Strasbourg, France.
}



\maketitle

\abstract{\textbf{Motivation:} With the growth of big data, variable selection has become one of the major challenges in statistics. Although many methods have been proposed in the literature their performance in terms of recall and precision are limited in a context where the number of variables by far exceeds the number of observations or in a high correlated setting.\\
\textbf{Results:} In this article, we propose a general algorithm which improves the precision of any existing variable selection method. This algorithm is based on highly intensive simulations and takes into account the correlation structure of the data. Our algorithm can either produce a confidence index for variable selection or it can be used in an experimental design planning perspective. We demonstrate the performance of our algorithm on both simulated and real data.\\
\textbf{Availability:} Code will be available as the \texttt{selectboost} package on the CRAN.\\
\textbf{Contact:} \href{i.aouadi@unistra.fr}{i.aouadi@unistra.fr} and \href{fbertran@math.unistra.fr}{fbertran@math.unistra.fr}\\
\textbf{Supplementary information:} Supplementary data are available at \textit{Bioinformatics}
online.}

\section{Introduction}

Technological innovations make it possible to measure large amounts of data in a single observation. As a consequence, problems in which the number $P$ of variables is greater that the number $N$ of observations have become common. As reviewed by Fan and Li \citep{fan2006}, such situations arise in many fields from fundamental sciences to social science, and variable selection is required to tackle these issues. For example, in biology/medicine, thousands of messenger RNA (mRNA) expressions \citep{lipshutz1999} may be potential predictors of some disease. Moreover, in such studies, the correlation between variables is often very strong \citep{segal2003} and variable selection methods often fail to make the distinction between the informative variables and those which are not. In this paper, we propose a general algorithm that enhances model selection in correlated variables.\\

First, we will assume a statistical model with a response variable $\mathbf{y} = (y_1,...,y_N)'$ (with the symbol "'" as the transposed), a variable matrix of size $N \times P$, $\mathbf{X}=(\mathbf{x}_{1.},...,\mathbf{x}_{P.})$ and a vector of parameters $\bm{\beta} = (\beta_1,...,\beta_P)'$.
Then, we will assume that the vector of parameters $\bm{\beta} = (\beta_1,...,\beta_P)'$ is sparse. In other words, we will assume that $\beta_i=0$ except for a quite small proportion of elements of the vector. We note $\mathcal{S}$ as the set of indices for which $\beta_i \neq 0$ and $q<\infty$ is the cardinality of this set $\mathcal{S}$. Without any loss of generality, we will assume that $\beta_p \neq 0$ if and only if $p \leq q$.\\

When dealing with a problem of variable selection, one of the goals is the estimation of the support, in which you want  $\mathbb{P}(\mathcal{S} = \mathcal{\hat{S}})$ to be close to one, with $\mathcal{\hat{S}}=\{k : \hat{\beta}_k\neq 0\}$. Here, our interest is mainly as follows, \textit{i.e.} in identifying the correct support $\mathcal{S}$. This kind of issue arises in many fields, for example in biology, where it is of greatest interest to discover which specific molecules are involved in a disease \citep{fan2006}.\\

There is a vast literature dealing with the problem of variable selection in both statistical and machine learning areas \citep{fan2006,fan2010}. The main variable selection methods can be gathered in the common framework of penalized likelihood. The estimate $\hat{\bm{\beta}}$ is then given by:

\begin{equation}\label{eq:01}
\hat{\bm{\beta}} = \underset{\bm{\beta} \in \mathbb{R}^P}{\arg\min} \left[- {\ell_N ({\bm{\beta}} ) }  +  \sum_{p=1}^P \pen_{\lambda}({\beta_p})\right],
\end{equation}

\noindent where $\ell_N (.)$ is the log-likelihood function, $\pen_{\lambda}(.)$ is a penalty function and $\lambda \in \mathbb{R}$ is the regularization parameter. As the goal is to obtain a sparse estimation of the vector of parameters $\bm{\beta}$, a natural choice for the penalty function is to use the so-called $\mathcal{L}_0$ norm ($\norm{.}_0$) which corresponds to the number of non-vanishing elements of a vector:

\begin{equation}\label{eq:02}
\begin{array}{lcccc}  
 \pen_{\lambda} &:&  \mathbb{R} & \mapsto &\{0,\lambda\}\\
    &&x &\mapsto&\left\{
\begin{array}{c l}     
   \pen_{\lambda} (x)= \lambda & \text{if } x\neq 0\\
   \pen_{\lambda} (x)= 0 &\text{else}
\end{array}\right.
\end{array}  
\end{equation}

\noindent which induces $\sum_{p=1}^P \pen_{\lambda}({\beta_p}) = \lambda \norm{\bm{\beta}}_0$. For example, when $\lambda = 1$, we get the Akaike Information Criterion (AIC) \citep{akaike1974} and when  $\lambda = \frac{\log(N)}{2}$ we get the Bayesian Information Criterion (BIC) \citep{schwartz1978}.\\ 

Many different penalties can be found in the literature. Solving this problem with  $\|.\|_0$ as part of the penalty is an NP-hard problem \citep{natarajan1995,fan2010}. It cannot be used in practice when $P$ becomes large, even when it is employed with some search strategy like forward regression, stepwise regression \citep{hocking1976}, genetic algorithms \citep{koza1999}. Donoho and Elad \citep{donoho2003} showed that relaxing  $\norm{.}_0$  to norm $\norm{.}_1$ ends, under some assumptions, to the same estimation. This result encourages the use of a wide range of penalties based on different norms. For example, the case where $\pen_{\lambda}({\beta_p}) = \lambda |\beta_p|$ is the Lasso estimator \citep{tibshirani1996} (or equivalently Basis Pursuit Denoising \citep{chen2001})  whereas $\pen_{\lambda}({\beta_p}) = \lambda \beta_p^2$ leads to the Ridge estimator \citep{hoerl1970}. Nevertheless, the penalty term induces variable selection only if:

\begin{equation}\label{eq:03}
\min_{x \geq 0}\left( \frac{\text{d}\pen_{\lambda}(x)}{\text{d}x}+x \right) >0.
\end{equation}

This explains why the Lasso regression allows for variable selection while the Ridge regression does not. The Lasso regression is however known to lead to a biased estimate \citep{zou2006}. The SCAD (Smoothly Clipped Absolute Deviation) \citep{fan1997}, MCP (Minimax Concave Penalty) \citep{zhang2010} or adaptive Lasso \citep{zou2006} penalties all address this problem. The popularity of such variable selection methods is linked to fast algorithms like LARS (Least-Angle Regression Selection) \citep{efron2004}, coordinate descent or PLUS (Penalized Linear Unbiased Selection) \citep{zhang2010}.\\

Nevertheless, the goal of identifying the correct support of the regression is complicated and the reason why variable selection methods fail to select the set of non-zero variables $\mathcal{S}$ can be summarized in two words: linear correlation. Choosing the Lasso regression as a special case, Zhao and Yu (2006) (and simultaneously Zou (2006)) found an almost necessary and sufficient condition for Lasso sign consistency (\textit{i.e.} selecting the  non-zero variables with the correct sign). This condition is known as "irrepresentable condition": 

\begin{equation}\label{eq:04}
\left| \mathbf{X}'_{\setminus\mathcal{S}}\mathbf{X}_{\mathcal{S}} \left(  \mathbf{X}'_{\mathcal{S}}\mathbf{X}_{\mathcal{S}}\right)^{-1}   \text{sgn}(\bm{\beta}_\mathcal{S})\right|< \bm{1},
\end{equation}

\noindent where $\mathbf{X}_\mathcal{S} = (x_{ij})_{i,j \in \mathcal{S}^2}$, $\mathbf{X}_{\setminus\mathcal{S}} = (x_{ij})_{i,j \neq \mathcal{S}^2}$, $\bm{\beta}_\mathcal{S} = ({\beta_p})_{p \in \mathcal{S}}$. In other words, when $ \text{sgn}(\bm{\beta}_\mathcal{S})=1$, this can be seen as the regression of each variable which is not in $\mathcal{S}$ over the variables which are in $\mathcal{S}$. As all variables in the matrix $\mathbf X$ are centered, the absolute sum of the regression parameters should be smaller than 1 to satisfy this "irrepresentable condition".\\

Facing this issue, existing variable selection methods can be split into two categories:

\begin{itemize}
\item those which are "regularized" and try to give similar coefficients to correlated variables (\textit{e.g.} elastic net \citep{zou2005}), 
\item those which are not "regularized" and pick up one variable among a set of correlated variables (\textit{e.g.} the Lasso  \citep{tibshirani1996}).
\end{itemize}

The former group can further be split into methods in which groups of correlation are known, such as the group Lasso \citep{yuan2006,friedman2010} and those in which groups are not known  as in  the elastic net \citep{zou2005}. The latter  combines the $\mathcal{L}_1$ and the $\mathcal{L}_2$ norm and takes advantage of both. Broadly speaking, non-regularized methods will select some co-variables among a group of correlated variables while regularized methods will select all variables in the same group with similar coefficients.\\
 
The main idea of our algorithm is to consider that groups of variables of the matrix $\mathbf{X}$ which are linearly correlated are independent realizations of the same random function. According to this random function, correlated variables are then perturbed.  Strictly speaking, the use of  noise to determine the informative variables  is not a new idea. For example, it has been shown that adding random pseudo-variables decreases over-fitting \citep{wu2007}. In the case where $P>N$ the pseudo-variables are generated either with a standard normal distribution $\mathcal{N}(0,1)$ or by using  permutations on the matrix $\mathbf{X}$ \citep{wu2007}. Another approach consists in adding noise to the response variable and leads to similar results \citep{luo2006}. The rational of this last method is based on the work of Cook and Stefanski \citep{cook1994} which introduces the  simulation-based algorithm SIMEX \citep{cook1994}. Adding noise to the matrix $\mathbf{X}$ has already been used in the context of microarrays \citep{chen2007}. Simsel \citep{eklund2012} is an algorithm that both adds noise to variables and uses random pseudo-variables. One new and interesting approach is stability selection \citep{meinshausen2010}  in which the variable selection method is applied on sub-samples, and informative variables are defined as variables which have a high probability of being selected. Bootstraping has been applied to the Lasso on both response variable and the matrix $\mathbf{X}$ with better results in the former case \citep{bach2008}. The random Lasso, in which variables are weighted with random weights, has also been introduced \citep{wang2011}.\\

In this article, following the idea of using simulation to enhance the variable selection methods, we propose the selectBoost algorithm. Unlike other algorithms reviewed above, it takes into account the correlation structure of the data. Furthermore, our algorithm is motivated by the fact that in the case of non-regularized variable selection methods, if a group contains variables that are highly correlated together, one of them will be chosen with precision.




\section{Methods}

The selectBoost algorithm has been designed in a general framework in order to avoid to select non-predictive correlated features. The main goal is to improve the precision, \textit{i.e.} the proportion of selected variables which truly belong to $\mathcal{S}$.\\

\subsection{Generate new perturbed design matrix}

As we assume that the variables are centered  and that $\norm{\mathbf{x}_p.}^2=1$ for $p=1,...,P$, we know that $\mathbf{x}_{p.} \in \mathcal{S}^{N-2}$. Indeed, the normalization puts the variables on the unit sphere $\mathcal{S}^{N-1}$. The process of centering can be seen as a projection on the hyperplane $\mathcal{H}^{N-1}$ with the unit vector as normal vector. Moreover, the intersection between $\mathcal{H}^{N-1}$ and $\mathcal{S}^{N-1}$ is $\mathcal{S}^{N-2}$. We further define the following isomorphism:

\begin{equation}\label{eq:06}
\begin{array}{cccccc}
\phi & : & \mathcal{H}^{N-1}   & \to & \mathbb{R}^{N-1} &\\
 & &  {\mathbf{h}_n}& \mapsto & \phi(\mathbf{h}_n)= \mathbf{f}_n & n=1,...,N-1,\\
\end{array}
\end{equation}

\noindent where  $\{ {\mathbf{h}_n}\}_{n=1,...,N-1}$ is an orthogonal base of $ \mathcal{H}^{N-1}$ and $\{\mathbf{f}_n \}_{n=1,...,N-1}$ is the canonical base of $\mathbb{R}^{N-1}$. We define:

$$
{\mathbf{h}_n} =\frac{ \sum_{i=1}^{n} \mathbf{e}_i - n \mathbf{e}_{n+1}}{\norm{\sum_{i=1}^{n} \mathbf{e}_i - n \mathbf{e}_{n+1}}},
$$

\noindent with $\{\mathbf{e}_n\}_{n=1,...,N} $ the canonical base of $\mathbb{R}^{N}$. Note that $\phi (\mathcal{S}^{N-2})=\mathcal{S}^{N-2}$, and that is why we can work in $ \mathbb{R}^{N-1}$ and then return in $ \mathbb{R}^{N}$.\\

Here, we make the assumption that a group of correlated variables are independent realizations of the same multivariate Gaussian distribution. As the variables are normalized with respect to the $\mathcal{L}_2$ norm, we will use the von Mises-Fisher distribution \citep{sra2012} in $\mathbb{R}^{N-1}$ thanks to the isomorphism $\phi$ in order to generate new perturbed design matrix. The probability density function of the von Mises-Fisher distribution for the random $P$-dimensional unit vector $\mathbf{x}\,$ is given by:

$$
f_{P}(\mathbf{x};\bm\mu, \kappa)=\widetilde{K}_{P}(\kappa)\exp \left( {\kappa \bm \mu' \mathbf{x} } \right),
$$

\noindent where $\kappa \geq 0$, $ \bm\mu =(\mu_1,...,\mu_P)'$,  $\norm{\bm\mu}_2 =1,$ and the normalization constant $\widetilde{K}_{P}(\kappa)$ is equal to:

$$
\widetilde{K}_{P}(\kappa)=\frac {\kappa^{P/2-1}} {(2\pi)^{P/2}I_{P/2-1}(\kappa)}, 
$$

\noindent where $I_{v}$ denotes the modified Bessel function of the first kind and order $v$ \citep{abramowitz1972}.

\subsection{The selectBoost algorithm}

To use the selectBoost algorithm, we need a grouping method $gr_{c_0}$ depending on a user-provided constant $0 \leq c_0 \leq 1$. This constant determines the strength of the grouping effect. The grouping method maps each variable index $1,...,P$ to an element of $\mathcal{P}(\{1,...,P\})$ (with $\mathcal{P}(S)$ the powerset of the set $S$, \textit{i.e.} the set which contains all the subsets of $S$). Concretely, $gr_{c_0}(p)$ is the set of all variables which are considered to be linked to the variable $\mathbf{x}_p$ and $\mathbf{X}_{gr_{c_0}(p)}$ is the submatrix of $\mathbf{X}$ containing the columns which indices are in $gr_{c_0}(p)$. We impose the following constraints to the grouping function:

\begin{equation}\label{eq:07}
\forall p \in \{1,...P\}: gr_1(p)=\{p\} \text{~~and~~} gr_0(p)=\{1,...P\}.
\end{equation}

Furthermore, we need to have a selection method: 
$$\begin{array}{ccccc} select & : & \mathbb{R}^{N \times P } \times \mathbb{R}^N& \to & \{0,1\}^P \end{array}$$ 

\noindent which maps the design matrix $\mathbf{X}$ and the response variable $\mathbf{y}$ to a 0-1 vector of length $P$ with $1$ at position $p$ if the method selects the variable $p$ and 0 otherwise.\\

\begin{algorithm}
\caption{Pseudo-code for the selectBoost algorithm}
\begin{algorithmic}
\REQUIRE $gr_{c_0},{select}, B,c_0$
\STATE $\bf\zeta \leftarrow\bf{0}_P$
\FOR{$b=1,...,B$}
\STATE $\bf{X}^{(b)} \leftarrow\bf{X}$
\FOR{$p=1,...,P$}
\STATE $\bf{x}^{(b)}_{p.} \leftarrow \phi^{-1} \left(\text{random-vMF}\left(\hat{\bf \mu}(\phi(\bf{X}_{gr_{c_0}(p)} )),\hat\kappa(\phi(\bf{X}_{gr_{c_0}(p)} )\right) \right)$
\ENDFOR
\STATE $\bf\zeta \leftarrow \bf\zeta + {select}(\bf{X^{(b)}},\bf y)$
\ENDFOR
\STATE $\bf\zeta \leftarrow \bf\zeta /B$
\end{algorithmic}
\end{algorithm}

We then use the von Mises-Fisher law to generate replacement of the original variables by some simulations (see Algorithm 1) to create $B$ new design matrices $\mathbf{X}^{(1)},...,\mathbf{X}^{(B)}$. The selectBoost algorithm then applies the variable selection method $select$ to each of these matrices and returns a vector of length $P$ with the frequency of apparition of each variable. The frequency of apparition of variable $\mathbf{x}_{p.}$, noted $\zeta_p$ is assumed to be  an estimator of the probability $\mathbb{P}(\mathbf{x}_{p.} \in \mathcal{S} ) $ for this variable to be in $\mathcal{S}$. The choice of $c_0$  are crucial. On the one hand, when this constant is too large, the model is not  perturbed enough. On the other hand, when this constant is too small, variables are chosen at random.\\

The selectBoost algorithm returns the vector $\bm\zeta = (\zeta_1,...,\zeta_P)'$. One has now to choose a threshold to determine which variables are selected. In this article, we choose to select a variable $p$ if $\zeta_p=1$. In some applications, lower choices of threshold may be chosen. 


\subsection{Choosing the parameters of the algorithm}

We first have to choose the grouping function. One of the simplest ways to define a grouping function $gr_{c_0}$ is the following:

\begin{equation}\label{eq:08}
gr_{c_0}(p) = \Big\{ q \in \{1,...,P\} \ \big| \ \mid <\mathbf{x}_{p.},\mathbf{x}_{q.}> \mid \ \geq c_0 \Big\} .
\end{equation}

In other words, the correlation group of the variable $p$ is determined by variables whose correlation with $\mathbf{x}_{p.}$ is at least $c_0$. In another way, the structure of correlation may further be taken into account using graph community clustering. Let $\bm C$ be the correlation matrix of matrix $\mathbf X$. Let define $\check{\bm C}$ as follows: 

$$
\check{c}_{ij}= \left\{\begin{array}{ccc}
\mid \check{c}_{ij} \mid &\text{if}& \mid \check{c}_{ij} \mid > c_0 \text{~~and~~} i\neq j\\
0& \text{otherwise.}&
\end{array}\right.
$$

Then, we apply a community clustering algorithm on the undirected network with weighted adjacency matrix defined by $\check{\bm C}$.\\

Once the grouping function is chosen we have to choose parameter $c_0$ Due to the constraints in equation (\ref{eq:07}) the selectBoost algorithm results in the initial variable selection method when $c_0=1$. As  we will show in the next section, the smaller the parameter $c_0$, the higher the precision of the resulting selected variables. On the other hand, it is obvious that the probability of choosing none of the variables (\textit{i.e.} resulting in the choice of an empty set) increases as the parameter $c_0$ decreases. In the perspective of experimental planning, the choice of $c_0$ should result of a compromise between  precision and  proportion of active identified variables. Hence, the $c_0$ parameter can be used to  introduce a confidence index $\gamma_p$ related to the variable $\mathbf{x}_{p.}$:

\begin{equation}
\gamma_p = 1 - \min_{\mathbf{x}_{p.} \in \hat{\mathcal{S}}_{c_0}} c_0, \textup{ hence } \ 0 \leq \gamma_p \leq 1
\label{eq:09}
\end{equation}

\subsection{Numerical studies}

In this section, we will assume a logistic model with a binary response variable \citep{peng2010}.\\

To assess the performance of the selectBoost algorithm, we performed indeed numerical studies. As stated before, the selectBoost algorithm can be applied to any existing variable selection method. Here, we decided to use the Lasso selection method. The performance of the Lasso method is known to be strongly dependent on the choice of the penalty parameter $\lambda$. In our simulations, we used a k-fold cross-validation to choose this penalty parameter. \\

To demonstrate the performance of the selectBoost method, we compared our method with stability selection \citep{meinshausen2010} and with a naive version of our algorithm, naiveSelectBoost. The naiveSelectBoost algorithm works as follows: estimate $\beta$ with any variable selection method then if $gr_{c_0}(p)$, as defined in equation (\ref{eq:08}) for example, is not reduced to {p}, shrink  to $0$. The naiveSelectBoost algorithm is similar to the selectBoost algorithm, except that it does not take into account the error which is made choosing at random a variable among a set of correlated variables.\\

We explored a situation repeated 100 times, the number of variables is 1000 and the number of observations is 100. In this situation, the response variable is linear but was transformed as a binary variable ($+1$ when $Y_i>0$ and $-1$ when $Y_i<0$) in order to analyze the logistic model. Data are generated from a cluster simulation \citep{bastien2014,bair2006}.  \\

\noindent
\textbf{Situation:} We are in a case where there is a linear link between the response and the only 50 first predictors and the last 950 variables are randomly generated from a standard normal distribution. \\

We use four indicators to evaluate the abilities of our method on simulated data. We define: 

\begin{itemize}
\item recall as the ratio of the number of correctly identified variables (\textit{i.e.} $\hat\beta_i \neq 0$ and $\beta_i \neq 0$) over the number of variables that should have been discovered (\textit{i.e.} $\beta_i \neq 0$).
\item precision as the ratio of correctly identified variables (\textit{i.e.} $\hat\beta_i \neq 0$ and $\beta_i \neq 0$) over the number of identified variables (\textit{i.e.} $\hat\beta_i \neq 0$).
\item Fscore as the following ratio: 
$$
2 \times \frac{\text{recall}\times \text{precision}}{\text{recall}+\text{precision} } \cdot
$$
\item selection as the average number of identified variables (\textit{i.e.} $\hat\beta_i \neq 0$).
\end{itemize} 

Note that our interest is focused on precision, as our goal is to select reliable variables. As stated before, when $c_0$ is decreasing toward zero, we expect a profit in precision and a decrease in recall. We also compute the Fscore which combines both recall and precision. As an improvement of precision comes with a decrease of the number of identified variables, the best method is the one with the highest precision for a given level of selection.


\begin{figure}[!b]
\centering
\includegraphics[height=6cm,width=8cm]{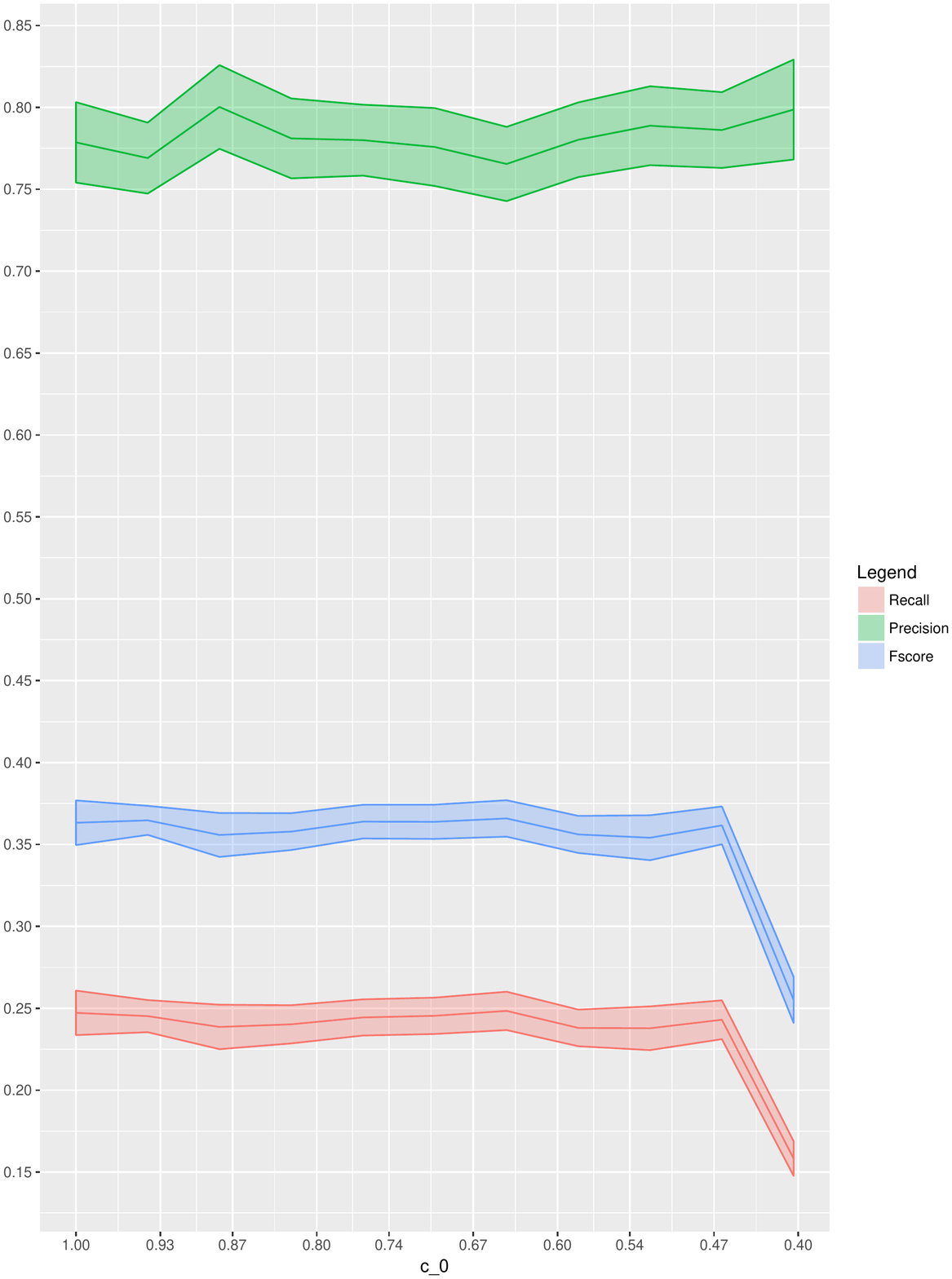}
\includegraphics[height=6cm,width=8cm]{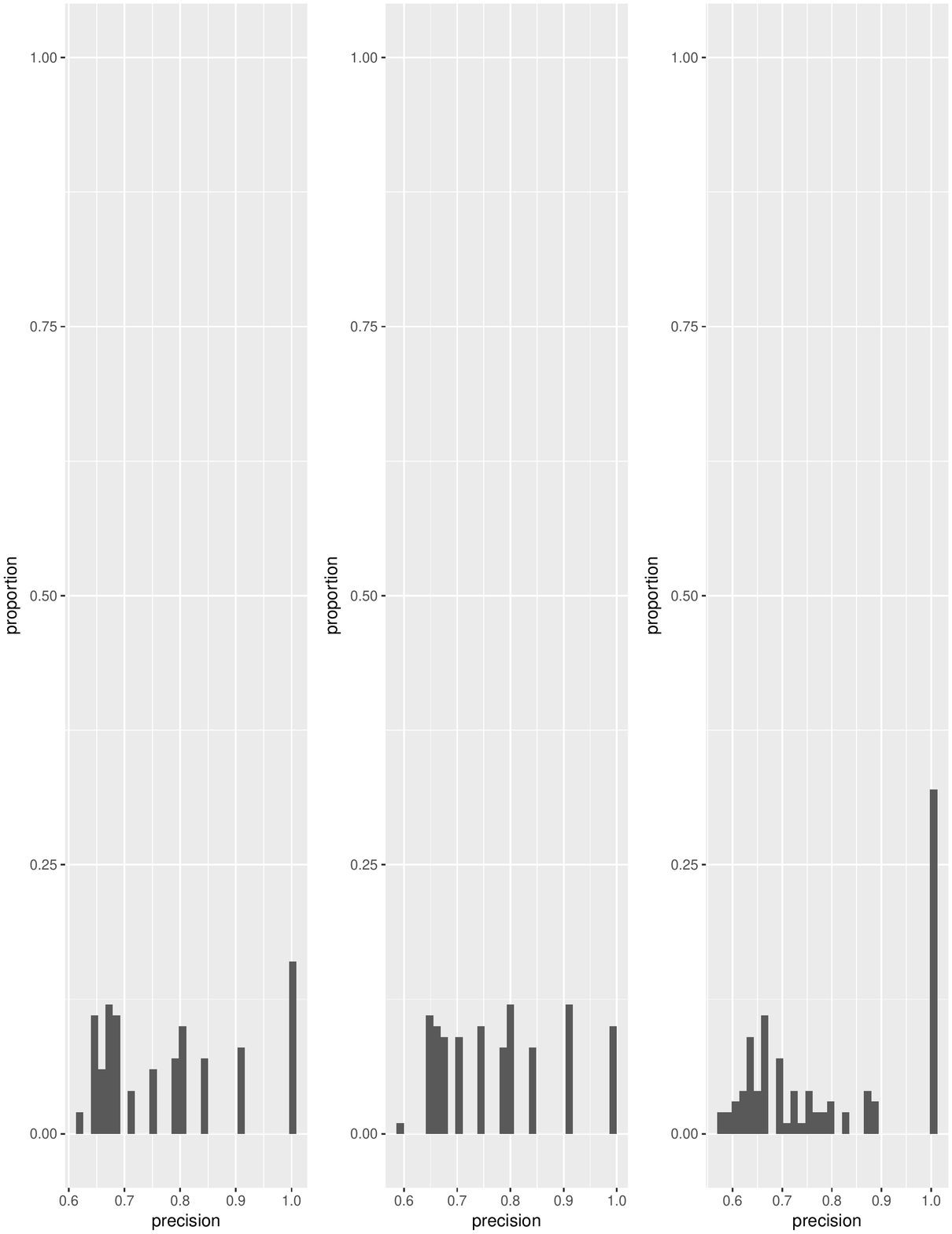}
\caption{Top: evolution of the recall, precision and Fscore in function of $c_0$. Bottom: the distribution of the precision among all models for the highest, an intermediate, and the lowest $c_0$.}
\label{ff}
\end{figure}

\section{Results of the numerical studies}

We first analyze the results for each situation. We show the evolution of the four criteria (recall, precision, Fscore and selection) with regards to the decrease of $c_0$. When $c_0=1$, the selectBoost algorithm is equivalent to the initial variable selection method, the Lasso method. As our main focus is precision, we add three histograms representing the evolution of the precision distribution for the highest, an intermediate and the lowest $c_0$. Figure \ref{ff} shows the result for the Lasso selection with penalty parameter chosen by cross-validation in Situation 1. In this example, we succeed in improving the precision up to 0.80. Moreover, as shown by the histograms of the precision, the proportion of models, for which the precision reaches one, increases with the decrease of $c_0$. The Fscore remains either stable or shows a small decrease indicating that the loss in recall is compensated by the increase of precision.\\
 
\begin{figure}[!tpb]
\centering
\includegraphics[height=7cm,width=7cm]{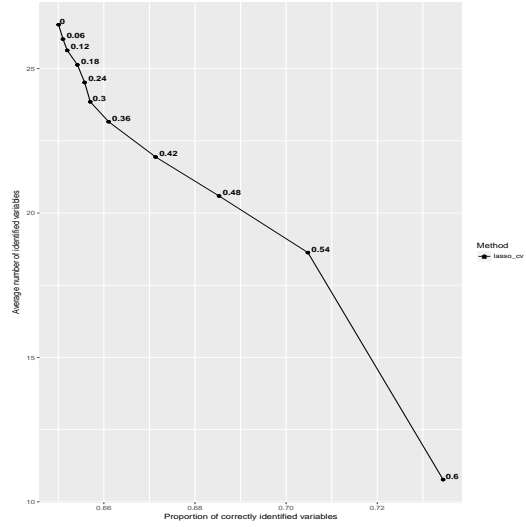}
\caption{The average number of identified variables is plotted in function of the proportion of correctly identified variables.}\label{ffff}
\end{figure}

In the previous section, we mentioned the possibility of using selectBoost to obtain a confidence index, corresponding to one minus the lowest $c_0$ for which a variable is selected. For each $c_0$, we plotted the average number of selected variables in function of the proportion of correctly identified variables (Figure \ref{ffff}). As expected, the proportion of correctly identified variables increases with the increase of the confidence index and with the decrease of the average number of identified variables. Therefore, the proportion of non-predictive features decreases with the increase of the confidence index. \\

The selectBoost algorithm shows its superiority over the naiveSelectBoost algorithm (Figure 4). The error which is made when choosing randomly a variable among a set of  correlated variables leads to further wrong choice of variables. While the intensive simulation of our algorithm allows to take into account this error, the naiveSelectBoost does not.\\

Finally we compare the selectBoost algorithm with stability selection(Figure 4). Stability selection use a re-sampling algorithm to determine which of the variables included in the model are robust. In our simulation, stability selection shows performance with high precision but also low proportion of recall. Moreover, in contrast to the selectBoost algorithm, stability selection does not allow to choose a convenient precision-recall trade-off.

\section{Application to a real dataset}

We decided to apply our algorithm to a real RNA-Seq dataset providing mRNA expressions from Huntington's disease and neurologically normal individuals. This dataset was downloaded from the GEO database under accession number GSE64810 (\url{https://www.ncbi.nlm.nih.gov/geo/query/acc.cgi?acc=GSE64810}). This dataset contains 20 Huntington's Disease cases and 49 neurologically normal controls, and includes 28,087 genes as explanatory variables.\\

We first applied the Lasso selection method to this dataset (see Figure \ref{huntington} left for the whole path of the solution). We used cross-validation to choose the appropriate level of penalization (\textit{i.e.} the $\lambda$ parameter in Equation (\ref{eq:03})). \\

\begin{figure}[!tpb]
\centering
\includegraphics[height=7cm,width=7cm]{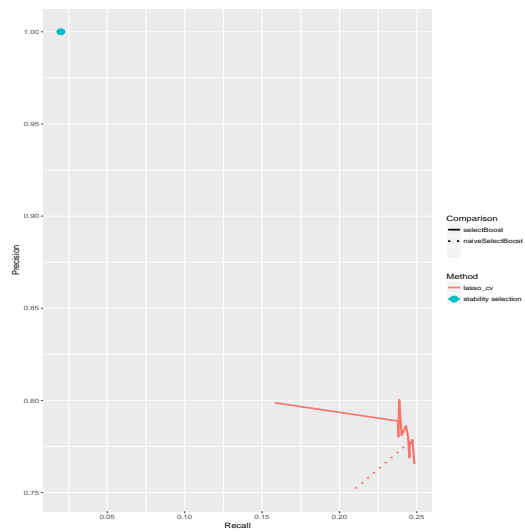}
\caption{Precision in function of recall. The selectBoost algorithm is compared to both stability selection and the naiveSelectBoost algorithm.}\label{fffff}
\end{figure}

We then applied our selectBoost algorithm on the Lasso method with penalty parameter chosen by cross-validation. We use a range for the $c_0$ parameter starting from 1 to 0.7 with steps of 0.05, which corresponds to a confidence index from 0 to 0.3. For each step, the probability of being included in the support $\mathcal{S}$ was calculated with 200 simulations as described in the Algorithm 1. We set the threshold of being in the support to 0.95 to avoid numerical instability. We classify the selected variables into three categories: those that are identified for each confidence index from $0$ to $0.15$ (red), those identified from $0$ to $0.25$ (orange) and those identified from $0$ to $0.3$ (green). The last category contains the most reliable variables selected by the selectBoost algorithm because these variables are identified from low to high confidence index.


With the Lasso selection method, 15 variables were selected. Among them, four genes were identified by selectBoost into the three different categories of confidence index (see Figure \ref{huntington} right): two genes for low confidence (red) (ANXA3 and INTS12), one gene for intermediate confidence (orange) (NUB1) and one gene for high confidence (green) (PUS3). \\




The interesting point, in these two examples, is that the identified variables are neither the first variables selected by the Lasso, nor the variables with the highest coefficients (see Figure \ref{huntington} left). This demonstrates that our algorithm can be very useful to select variables with high confidence and not just to select variables with the highest coefficients. \\

Finally, we decided to assess differential expression of these genes between patients and controls, using limma package (Linear Models for Microarray and RNA-Seq Data) \citep{ritchie2015}. The four identified genes are significantly down-expressed by neurologically normal controls, confirming the result of a logistic model including these four genes.
 

\section{Conclusion}

We introduce the selectBoost algorithm which uses intensive computation to select variables with high precision. The user of selectBoost can use this algorithm to produce a confidence index, or choose an appropriate precision-selection trade-off to select variables with high confidence and avoid selecting non-predictive features. The main idea behind our algorithm is to take into account the correlation structure of the data and thus use intensive computation to select reliable variables. We prove the performance of our algorithm through simulation studies in various settings. Indeed, we succeeded in improving the precision of Lasso selection method with a relative stability on recall and Fscore. Our results open the perspective of a precision-selection trade-off which may be very useful in some situations where many regressions have to be made (\textit{e.g.} in network reverse-engineering in which we have one regression per vertex). In such a context our algorithm may be used in an experimental design approach because we may define the strength of the correlation between variables. The application to a real dataset allowed us to show that the most reliable variables are not necessarily those with the highest coefficient. The selectBoost algorithm is a powerful tool that can be used in every situation where reliable and robust variable selection has to be made. 

\begin{figure}[!t]
\centering
\includegraphics[height=7cm,width=7cm]{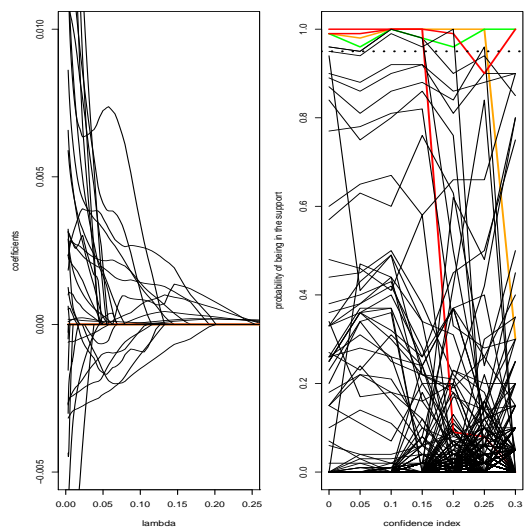}
\caption{\textbf{Colors:} the green is for the most reliable variables selected by the selectBoost algorithm (confidence index of 0.3; orange is for intermediate confidence (0.25) and red for low confidence (0.15)). \textbf{Left:} evolution of the coefficients in the Lasso regression when the regularization parameter $\lambda$ is varying. \textbf{Right:} evolution of the probability of being in the support of the regression when the confidence index is varying. The dotted line represents the threshold of 0.95.}
\label{huntington}
\end{figure}

\section*{Acknowledgements}

We are grateful to David Brusson from the Mesocenter of the University of Strasbourg.
%

\section*{Funding}

This work was supported by grants from the Agence Nationale de la Recherche (ANR) (ANR-11-LABX-0070\_TRANSPLANTEX), the INSERM (UMR\_S 1109), the Institut Universitaire de France (IUF), and the MSD-Avenir grant AUTOGEN, all to SB; the European regional development fund (European Union) INTERREG V program (project number 3.2 TRIDIAG) to RC and SB, as well as LabEx IRMIA to FB and MMB.\vspace*{-12pt}

%
%

\end{document}